\documentclass[twocolumn,prl,showpacs,preprintnumbers,amsmath,amssymb]{revtex4}
\usepackage{graphicx}
\usepackage{dcolumn}
\usepackage{bm}

\begin{document}

\preprint{APS/123-QED}
\title{Selfsynchronization and dissipation-induced threshold in collective atomic recoil lasing}

\author{C. von Cube}
\author{S. Slama}
\author{D. Kruse}
\author{C. Zimmermann}
\author{Ph.W. Courteille}
\affiliation{Physikalisches Institut, Eberhard-Karls-Universit\"at T\"ubingen,
\\Auf der Morgenstelle 14, D-72076 T\"ubingen, Germany}
\author{G.R.M.\ Robb$^{a}$, N.\ Piovella$^{b}$, and R. Bonifacio$^{b}$}
\affiliation{$^{a}$Department of Physics, University of Strathclyde, Glasgow, G4 0NG, Scotland.\\
$^{b}$Dipartimento di Fisica, Universit\`a Degli Studi di Milano and INFM,Via Celoria 16, I-20133 Milano, Italy.}

\date{\today}

\begin{abstract}
Networks of globally coupled oscillators exhibit phase transitions from incoherent to coherent states. Atoms interacting with the
counterpropagating modes of a unidirectionally pumped high-finesse ring cavity form such a globally coupled network. The coupling
mechanism is provided by collective atomic recoil lasing (CARL), i.e. cooperative Bragg scattering of laser light at an atomic
density grating, which is self-induced by the laser light. Under the rule of an additional friction force, the atomic ensemble is
expected to undergo a phase transition to a state of synchronized atomic motion. We present the experimental investigation of
this phase transition by studying the threshold behavior of the CARL process.
\end{abstract}

\pacs{42.50.Vk, 05.45.Xt, 05.65+b, 05.70.Fh}

\maketitle

Lasing mediated by collective atomic recoil (CARL) has been predicted as the analogue to free-electron lasing \cite{Bonifacio94}
and observed recently \cite{Kruse03b}. In the experiment an ensemble of cold atoms couples to the two counterpropagating modes of
a unidirectionally pumped high-finesse ring-cavity and collectively scatters photons between the modes. A collective instability
leads to self-amplification and to exponential gain for a light mode on one hand and atomic bunching on the other. In the absence
of energy dissipation for the kinetic degrees of freedom, the gain must remain transient. The introduction of a friction force
for the kinetic energy of the atoms permits however a steady-state operation of the CARL at a self-determined frequency (viscous
CARL) \cite{Note03}.

The CARL represents a system of coupled oscillators. Collective dynamics in networks of weakly coupled systems is a very general
phenomenon. The richness of this paradigm is illustrated by the large amount of examples, ranging from physical systems, like
arrays of Josephson junctions or phase-locked lasers, to biological systems like cardiac pacemaker cells or chorusing crickets
\cite{Strogatz01}. Kuramoto, Strogatz and others \cite{Kuramoto84,Strogatz00} have shown that ensembles of coupled oscillators
operating at different frequencies or being subject to stochastic noise synchronize, if their number and their mutual coupling
strength exceeds a critical value. I.e. cooperative action starts beyond a threshold value of the coupling strength after
crossing a thermodynamical phase transition.

In the present paper, we investigate the threshold behavior of the CARL as an example of a phase transition of the Kuramoto type
\cite{Kuramoto84,Strogatz00}: If, for viscous CARL, the dissipation (or cooling) mechanism is limited to finite temperatures by
some diffusion process, a phase transition occurs in the atomic density distribution when the system is pumped at threshold.
Consequently, a minimum pump power is necessary to start CARL lasing, just like in ordinary lasers. However, in extension to the
original Kuramoto model, the collective frequency is self-determined. We report on the observation of such a threshold and
characterize it in terms of the theoretical model presented in Ref.~\cite{Robb04}.

The optical layout of our experiment has been outlined in Ref.~\cite{Kruse03b}. A titanium-sapphire laser is tightly phase-locked
to one of the two counterpropagating modes of a ring cavity by means of a Pound-Drever-Hall type servo control. The amplitude
decay rate of the ring cavity is $\kappa=(2\pi)22$~kHz. In the following, we label the modes by their complex field amplitudes
scaled to the field per photon $\alpha_{\pm}$. The intracavity light power is then $P_{\pm}=\hbar\omega\delta|\alpha_{\pm}|^2$,
where $\delta=3.5~$GHz denotes the free spectral range of the cavity. The phase dynamics of the two counterpropagating cavity
modes is monitored via the beat signal between the two outcoupled beams: Any displacement of the standing wave inside the ring
cavity is translated into an amplitude variation of the observed interference signal, $P_{beat}=\hbar\omega\delta
|\alpha_++\alpha_-|^2$.

The contrast of the standing wave $P_{beat}$ is weak as compared to the pump power (a few \%). Assuming $\alpha_+$ real and
$|\alpha_-|\ll\alpha_+$, the probe beam power can be related to the contrast of the beat signal $\Delta P_{cont}
=4\hbar\omega\delta\alpha_+|\alpha_-|$,
\begin{equation}
P_-=\frac{\Delta P_{cont}^2}{16P_+}~.\label{EqContrast}
\end{equation}

In our experiment, we load $^{85}$Rb atoms from a standard magneto-optical trap (MOT) into the optical dipole potential, which is
generated by a TEM$_{00}$ mode of the unidirectionally pumped ring cavity and tuned to the red of the $D_1$ line. Typically
$2\times10^6$ atoms are trapped and form a basically homogeneous $4~$mm long cloud along the cavity axis around the waist of the
cavity field. The cloud reaches a peak density of about $2\times10^9$~cm$^{-3}$ and a temperature of a few $100~\mu$K. Absorption
images of the atomic density distribution taken after a time-of-flight and spectra of the velocity distribution obtained by
exciting recoil-induced resonances (RIR) \cite{Kruse03} yield information on the kinetic degrees of freedom of our system, which
is complementary to that on the field amplitudes $\alpha_\pm$.

A genuine problem of the CARL is the following: In the absence of damping for the external degrees of freedom the CARL process
continuously accelerates the atomic center-of-mass \cite{Perrin02b,Kruse03b}. Even though the acceleration decreases because the
Doppler-shifted CARL frequency eventually drops out of the cavity resonance, it never reaches a stationary value. In fact, being
focussed on studies of transient phenomena, the original CARL model \cite{Bonifacio94} does not consider relaxation of the
translational degrees of freedom. On the other hand, standard methods of optically cooling atoms are based on controlled
dissipation, e.g. optical molasses. Close to resonance the motion of atoms in an optical molasses is well described by a friction
force. In our experiment, we harness this dissipation mechanism and subject the dipole-trapped atomic cloud to an optical
molasses. We use the laser beams of the MOT and tune them $50~$MHz below the cooling transition ($D_2$, $F=3 \rightarrow F'=4$).
In this situation, the beat frequency oscillations quickly reach a stable equilibrium frequency between $\Delta\omega/2\pi=100~$kHz
and $170~$kHz, which corresponds to an atomic velocity of $7$ to $13~$cm/s.

\begin{figure}[ht]
  \centerline{\scalebox{0.42}{\includegraphics{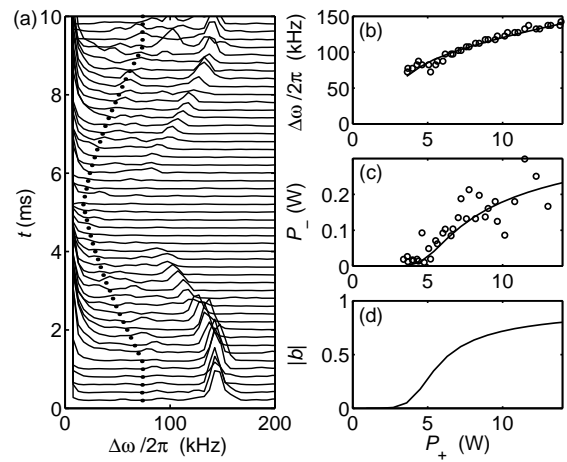}}}\caption{
  \textbf{(a)} Sectionwise Fourier-transform of the interference signal $P_{beat}$ with the pump laser power being ramped up
  and down. The dotted line is proportional to the pump laser power. \textbf{(b)} Dependence of the CARL frequency on the
  intracavity pump power. \textbf{(c)} Dependence of the probe field intensity on the pump power. The CARL laser threshold is around
  $P_+=4$~W intracavity power. The fitted curves are based on a Fokker-Planck theory outlined in \cite{Robb04}. \textbf{(d)}
  Calculated bunching parameter. The parameters are $\gamma_{fr}=4\kappa$, $N=10^6$, $\Delta_a=-(2\pi)1.7~$THz, and $T=200~\mu K$.}
  \label{CoopFigRamp}
  \end{figure}

In order to observe a threshold behavior in experiment, we adiabatically ramp up and down the intensity of the pump laser. The
CARL radiation is monitored by recording the time evolution of the beat frequency of the counterpropagating modes. The curves
shown in Fig.~\ref{CoopFigRamp}(a) represent frequency spectra obtained by Fourier transforming the beat signal restricted to
successive time-intervals. The peaks' locations then denote the instantaneous frequency shift of the probe beam, and their
heights reflect the standing wave's contrast. The pronounced dependence of the CARL frequency on the pump intensity revealed by
the series of Fourier spectra is emphasized in Fig.~\ref{CoopFigRamp}(b). The intensity of the CARL radiation shown in
Fig.~\ref{CoopFigRamp}(c) decreases with the pump intensity, but more important is the fact that the curve exhibits a minimum
pump intensity required to initiate CARL lasing. Just like in common lasers, only if the energy fed to the system exceeds the
losses, the laser emits coherent radiation.

The experimental observations may be discussed in various ways. The model of Ref.~\cite{Kruse03b} to describe the impact of
optical molasses on CARL consisted of simply adding a friction force, proportional to a coefficient $\gamma_{fr}$, to the
equations governing the atomic dynamics. This procedure certainly represents a coarse simplification. For example it predicts
that the atoms quickly bunch under the influence of the molasses, and are cooled until the temperature of the cloud is $T=0$, and
it denies the presence of any threshold. In reality, the molasses temperature is limited by diffusion in momentum space, i.e.
heating. To account for this heating, one may supplement the dynamic equations for the trajectories of individual atoms
(Ref.~\cite{Kruse03b}, Eq.~(1)) with a stochastically fluctuating Langevin force $\xi_n(t)$ with $\langle\xi_n(t)\rangle=0$ and
$\langle\xi_n(t)\xi_m(\tau)\rangle=2\gamma_{fr}^2 D\delta_{mn}\delta(t-\tau)$, where the diffusion coefficient
$D=\sigma^2/\gamma_{fr}$ is proportional to the atoms' equilibrium temperature, which is related to the Maxwell-Gaussian velocity
spread by $\sigma\equiv 2k\sqrt{k_BT/m}$:
\begin{equation}
\ddot{\theta}_n = 4\varepsilon iU_0\alpha_+\left(\alpha_-e^{-i\theta_n}-\alpha_-^*e^{i\theta_n}\right)-\gamma_{fr}\dot{\theta_n}
+\xi_n~.\label{EqAtom}
\end{equation}
Here we defined $\theta_n\equiv2kx_n$ as the position of the $n^{th}$ atom along the optical axis normalized to the optical
wavelength and assumed the pump laser stabilized on resonance with the cavity, $\alpha_+$ is set real and constant. $N$ is the
total atom number, and $\varepsilon\equiv\hbar k^2/m$ is twice the recoil frequency shift. The coupling strength $U_0$ (or
single-photon light shift) is related to the one-photon Rabi-frequency $g$ and to the laser detuning from resonance by $U_0\equiv
g^2/\Delta_a$. The functional dependence of the quantity $\alpha_-$ on the order (or bunching) parameter
$b\equiv|b|e^{i\psi}\equiv N^{-1}\sum\nolimits_me^{i\theta_m}$ is determined by the differential equation
\begin{equation}
\dot{\alpha}_- = -\kappa\alpha_--iNU_0\alpha_+~b~.\label{EqField}
\end{equation}
An alternative to simulating trajectories of individual atoms is to calculate the dynamics of distribution functions.
Particularly adequate to the problem of diffusion in momentum space induced by optical molasses is a Fokker-Planck approach. Here
the thermalization of the atomic \emph{density} distribution $P$ towards an equilibrium between cooling and heating is described
by the interplay of friction and diffusion \cite{Robb04}. In the limit of strong viscous damping, where we may adiabatically
eliminate the atomic momenta by setting $\ddot{\theta}_n=0$, the Fokker-Planck equation associated to the Langevin
equation~(\ref{EqAtom}) reads:
\begin{equation}
\frac{\partial P}{\partial t} = \frac{4i\varepsilon U_0\alpha_+}{\gamma_{th}}\frac{\partial\left(\alpha_-^{\ast}e^{2i\theta}
-\alpha_-e^{-2i\theta}\right)P}{\partial\theta}+D\frac{\partial^2 P}{\partial\theta^2}~,\label{EqFokker}
\end{equation}
and the bunching parameter is given by $b=\int_0^\infty Pe^{i\theta} d\theta$. The solid lines fitted to the data in
Fig.~\ref{CoopFigRamp} are calculated by numerical integration of this Fokker-Planck equation. Fig.~\ref{CoopFigRamp}(d) shows
the calculated evolution of the \emph{bunching}. Apparently, the bunching vanishes below the CARL lasing threshold and tends
towards $1$ as the pump power is increased. The threshold behavior of the radiation mode is thus intrinsically connected to
atomic self-organization. As has already been realized by Kuramoto, the Fokker-Planck equation predicts the occurrence of a
thermodynamical phase transition. An alternative approach developed by Bonifacio and Verkerk \cite{Bonifacio96} studies the
evolution of the atomic \emph{phase-space} distribution described by the Vlasov equation towards equilibrium with a single rate
$\gamma_{fr}$, and Javaloyes et al. \cite{Javaloyes04} found that the Vlasov approach leads to a phase transition.

Ref.~\cite{Robb04} pointed out that the threshold should depend on various parameters like the atom number, the coupling
strength, the friction coefficient and also on the atomic temperature. However a proper scaling \cite{Robb04} shows that the
threshold is ruled by only two independent quantities. Analytic expressions for the threshold conditions are obtained from a
linear stability analysis of the Fokker-Planck equations. As shown in Ref.~\cite{Robb04}, by introducing the CARL parameter
$\rho\equiv(NU_0^2\alpha_+^2 /2\varepsilon^2)^{1/3}$, the condition for lasing is given by
\begin{equation}
\frac{\kappa\sigma^2}{\gamma_{fr}}\left(\sigma^2+\kappa\gamma_{fr}\right)^2\leq\left(2\varepsilon\rho\right)^6~.\label{EqThresh}
\end{equation}
In the good cavity limit $\kappa\gamma_{fr}\leq\sigma^2$, we may neglect the second term in the bracket. At threshold, where the
equality in the above equation holds, the CARL parameter becomes $\rho_{thr}=\left(\kappa/\gamma_{fr}\right)^{1/6}
\left(\sigma/2\varepsilon\right)$. Therefore the threshold pump power $P_{thr}=\hbar\omega\delta\alpha_{+,thr}^2$ follows with
\begin{equation}
\alpha_{+,thr}^2=\left(\frac{\kappa}{\gamma_{fr}}\right)^{1/2} \frac{\sigma^3}{4\varepsilon NU_0^2}~.\label{EqPump}
\end{equation}
The threshold power and the threshold CARL frequency are related by \cite{Robb04}, $\Delta\omega_{thr}^3
=\left(2\varepsilon\rho_{thr}\right)^3 \kappa/\gamma_{fr}$. As a consequence, the CARL frequency is independent of the atom-field
coupling,
\begin{equation}
\Delta\omega_{thr}=\sigma\left(\frac{\kappa}{\gamma_{fr}}\right)^{1/2}~.\label{EqFreq}
\end{equation}

  \begin{figure}[h]
  \centerline{\scalebox{0.42}{\includegraphics{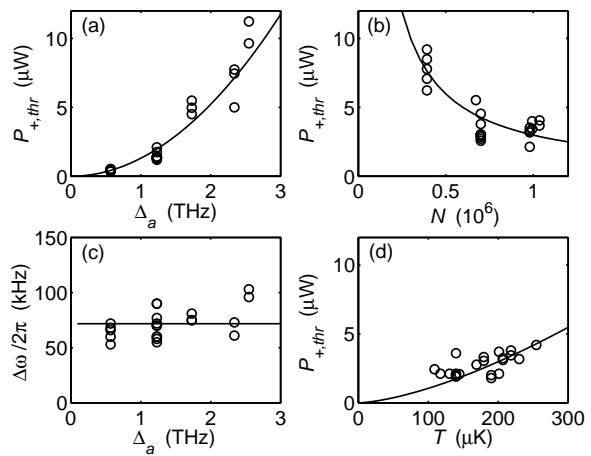}}}\caption{
  Pump power threshold measured for various laser detunings \textbf{(a)} and atom numbers \textbf{(b)} determined from curves like
  those shown in Fig.~\ref{CoopFigRamp}. \textbf{(c)} CARL frequency shifts as a function of the laser detuning. \textbf{(d)} Pump
  power threshold plotted as a function of the temperature derived from the threshold CARL frequency according to Eq.~(\ref{EqFreq}).
  }\label{CoopFigThreshold}
  \end{figure}

Figs.~\ref{CoopFigThreshold}(a) and (b) show measurements of the threshold pump power as a function of the laser detuning and the
total atom number. Both experimental curves have been obtained from the same data set $P_{+,thr}(N,\Delta_a)$ scaled either to an
arbitrarily chosen reference atom number $N=10^6$ (curve~(a)) or to a reference laser detuning $\Delta_a=-(2\pi)1.5~$THz
(curve~(b)) using the relationship (\ref{EqPump}). The fitted curves are obtained from the same equation (\ref{EqPump}) by
adjusting the temperature to $T=200~\mu$K and the friction coefficient to $\gamma_{fr}\approx 4\kappa$.
Fig.~\ref{CoopFigThreshold}(c) demonstrates that the CARL frequencies measured at threshold do apparently not depend on the
coupling strength. The horizontal line indicates the frequency corresponding to the temperature $T=200~\mu$K according to
Eq.~(\ref{EqFreq}). On the other hand, theory predicts a dependence of the threshold power on the atomic temperature: Hotter
atomic clouds have a broader velocity distribution $\sigma$ and exhibit a higher threshold power. Even though the temperature
data shown in Fig.~\ref{CoopFigThreshold}(d) seem to confirm this trend, they are too uncertain to be used to improve the fits of
Figs.~\ref{CoopFigThreshold}(a) and (b).
%In this respect the development of a reliable independent temperature measurement tool, e.g. by spectroscopy of recoil-induced
%resonances (RIR), seems desirable.
%  \begin{figure}[h]
%  \centerline{\scalebox{0.49}{\includegraphics{CoopFigThreshold.eps}}}\caption{
%  \textbf{(a)} Pump power threshold and \textbf{(b)} CARL frequency shift measured as shown in Fig.~\ref{CoopFigRamp} as a function
%  of the normalized coupling strength $NU_0^2/\kappa^2$.}\label{CoopFigThreshold}
%  \end{figure}

%Fig.~\ref{CoopFigThreshold}(a) shows measurements of the threshold pump power as a function of the total atom number $N$ and the
%single-atom coupling strength $U_0$. The fitted curve is obtained from Eq.~(\ref{EqPump}) by adjusting the friction coefficient
%to $\gamma_{fr}\approx 4\kappa$. Fig.~\ref{CoopFigThreshold}(b) shows the CARL frequencies measured at threshold. The horizontal
%line indicates the CARL frequency which corresponds to the temperature $T=200~\mu$K according to Eq.~(\ref{EqFreq}).

\begin{figure}[h]
  \centerline{\scalebox{0.42}{\includegraphics{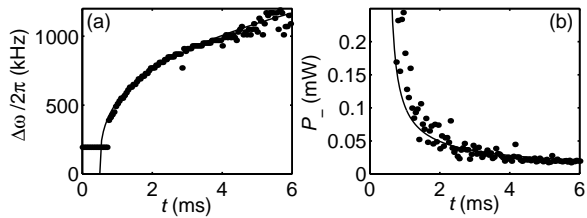}}}\caption{
  \textbf{(a)} Evolution of the CARL frequency after molasses has been switched off at time $t=0.5~$ms. \textbf{(b)} CARL power
  calculated via Eq.~(\ref{EqContrast}). The fits are based on the theoretical formulae Eqs.~(2) and (3) of Ref.~\cite{Kruse03b}.
  }\label{CoopFigAccelerate}
  \end{figure}

The steady-state operation of our system is ensured by the optical molasses. As soon as the molasses is turned off, the
equilibrium is lost and the atoms and the standing-wave accelerate each other, \emph{provided the atoms are bunched}.
Fig.~\ref{CoopFigAccelerate}(a) demonstrates the acceleration process. The instantaneous CARL frequency and intensity are again
obtained from a sequence of Fourier spectra taken over successive time periods. The evolution of the probe power calculated with
Eq.~(\ref{EqContrast}) is shown in Fig.~\ref{CoopFigAccelerate}(b). At times $t<0.5~$ms, when the molasses is present, the CARL
frequency is fixed at $170~$kHz. As soon as the molasses is switched off, the probe's amplitude diminishes while its frequency
detuning from the cavity resonance increases. The acceleration indicates the occurrence of bunching induced by the optical
molasses. The self-organization of the atomic density distribution out of a homogeneous cloud into a spatially ordered
arrangement spontaneously breaks translational symmetry and represents the strongest signature of CARL action \cite{Bonifacio94}.
%Note that the nature of the phase transition was the subject of thorough theoretical investigations in the past. Earlier CARL
%experiments \cite{Lippi96,Hemmer96} operated with room temperature cells in a regime, where the pump laser frequency was tuned
%within the Doppler-profile of the atomic resonance. In this case or in the presence of collisions, CARL action must not necessarily
%lead to a \emph{density} grating, and a certain synchronization of internal state polarization gratings with the atomic
%distribution in space is sufficient to give rise to a phase transition \cite{Perrin02b}. In contrast to those experiments, our
%pump laser has THz detuning and our rate of collisions is negligible, so that we do not expect a noticeable effect from atomic
%polarization gratings, which do not correlate to density gratings.

The probe field $\alpha_-$ depends on the location of all atoms in a collective way. Consequently, Eq.~(\ref{EqAtom}) describes a
\emph{mean-field} type dependence of the location of every single atom on all the others. This fact gets particularly transparent
in the limit of strong viscous damping, $\ddot{\theta_n}=0$, if we furthermore assume that at steady state the optical standing
wave propagates with a constant amplitude at a constant velocity. This condition is formulated by $\dot{\alpha}_-
=i\omega_{ca}\alpha_-$, where the probe beam frequency shift $\omega_{ca}=\omega_{ca}(b)$ depends on the atomic bunching. With
these approximations we substitute the solution of Eq.~(\ref{EqField}) into Eq.~(\ref{EqAtom}) and obtain, defining the coupling
constant $K\equiv 8\varepsilon NU_0^2\alpha_+^2\omega_{ca}/\left(\gamma_{fr}(\omega_{ca}^2+\kappa^2)\right)$ and restricting
ourselves to the good cavity limit $\kappa\ll\omega_{ca}$.
\begin{equation}
\dot{\theta}_n=\frac{\xi_n}{\gamma_{fr}}+K|b|\sin\left(\psi-\theta_n\right)~.\label{EqKuram}
\end{equation}
The equation (\ref{EqKuram}) describes the dynamics of $N$ coupled oscillators with fictitious frequencies
$\omega_n=\xi_n/\gamma_{fr}$. This system, which has been investigated by Kuramoto \cite{Strogatz00}, predicts the
synchronization of those oscillators over time, whose frequencies satisfy $\omega_n\leq K|b|$. The analogy between our viscous
CARL and an ensemble of self-synchronizing harmonic oscillators resides in the following correspondences: The phases of the
oscillators are represented by the positions of atoms. Synchronization of the oscillators corresponds to bunching of the atoms.
The role of friction is to provide a steady atomic center-of-mass velocity to which the individual atomic velocities may lock. In
the case of CARL and unlike for the Kuramoto model the collective oscillation frequency is self-determined. Diffusion is the
source of disorder, which rules the phase transition by competing with the dynamical coupling, in contrast to the Kuramoto model,
where disorder occurs via distributed natural frequencies.

To conclude we point out, that the viscous CARL system is representative for a vast class of systems. Under the rule of optical
molasses the coupled field-atom system constitutes an ideal model system for an ensemble of weakly coupled oscillators. Despite
the fact that the details of the molasses dynamics are complicated, its impact on our CARL system is fairly well described by two
constants, the friction and the diffusion coefficient. Although the system is purely classical, we deal with microscopic
particles. This bears the possibility of transferring the system to the quantum regime, and thus to study the coupling of large
ensembles of quantum oscillators. Furthermore, classical networks of dynamical systems in general depend much on the details of
how the coupling is realized. In contrast, the coupling in our system is generated by the fundamental interaction between atoms
and light, which is very well understood and even controllable by experiment, e.g. via the tunable friction force or the laser
detuning. Because it is mediated by a delocalized object, i.e. a standing light wave, the coupling is instantaneous (no
retardation effects) and truly uniform (every atom couples with the same strength to all its neighbors). The large number of
oscillators ensures that the system is in the thermodynamic limit.

%As stated in Ref.~\cite{Robb04}, in the bad cavity limit, steady-state superradiance should be possible. Note that in this limit
%the functional dependence of the CARL effect on the atom number is different: A quadratic dependence of the reverse field
%intensity on the atom number would be an indication for superradiant scattering.
%A major challenge for future experiments remains the \emph{direct} detection of the spatial ordering of the atoms as a
%consequence of CARL action. This could be done using Bragg scattering techniques \cite{Weidemuller95}.

\bigskip

We are grateful for valuable discussions with J. Javaloyes who brought the Kuramoto model to our attention. We acknowledge
financial support from the Landesstiftung Baden-W\"urttemberg. GRMR, NP and RB acknowledge support from the Royal Society.

\end{document}